# Exploring Human Mobility Patterns Based on Location Information of US Flights

Bin Jiang and Tao Jia

Department of Technology and Built Environment, Division of Geomatics
University of Gävle, SE-801 76 Gävle, Sweden
Email: bin.jiang@hig.se, jiatao83@163.com

*(Draft: April 2011, Revision: August 2011)*

**Abstract**
A range of early studies have been conducted to illustrate human mobility patterns using different tracking data, such as dollar notes, cell phones and taxicabs. Here, we explore human mobility patterns based on massive tracking data of US flights. Both topological and geometric properties are examined in detail. We found that topological properties, such as traffic volume (between airports) and degree of connectivity (of individual airports), including both in- and outdegrees, follow a power law distribution but not a geometric property like travel lengths. The travel lengths exhibit an exponential distribution rather than a power law with an exponential cutoff as previous studies illustrated. We further simulated human mobility on the established topologies of airports with various moving behaviors and found that the mobility patterns are mainly attributed to the underlying binary topology of airports and have little to do with other factors, such as moving behaviors and geometric distances. Apart from the above findings, this study adopts the head/tail division rule, which is regularity behind any heavy-tailed distribution for extracting individual airports. The adoption of this rule for data processing constitutes another major contribution of this paper.

**Introduction**
Human mobility is a research topic of primary interest to many disciplines, such as geography, urban planning, epidemiology and even telecommunication (Hägerstrand 1970, Hillier et al. 1993, Hufnagel, Brockmann and Geisel 2004, Lee et al. 2009). Recently, with the availability of massive amounts of various tracking data, researchers have attempted to investigate the underlying regularities and mechanisms of human movement patterns (Brockmann, Hufnage and Geisel 2006, Gonzalez, Hidalgo and Barabási 2008, Jiang, Yin and Zhao 2009, Song et al. 2010, Roth et al. 2011). It was found that human mobility exhibits a scaling property with a high probability of predicting individual and collective movement patterns. The tracking data related to US dollar notes (Brockmann, Hufnage and Geisel 2006), cell phones (Gonzalez, Hidalgo and Barabási 2008, Song et al. 2010), and taxicabs (Jiang, Yin and Zhao 2009) have been used. The scaling property indicates that human activities in geographic space are neither random nor even but demonstrate a very high degree of heterogeneity in terms of distances and preferred places. In this paper, we adopt location information of US flights, captured every 5 minutes by GPS receivers while the planes are in service. In total, over 7 million locations of various US flights were used to extract information related to both geometric and topological properties, such as travel lengths, traffic volume, and airport degrees of connectivity (by routes or by flights) for further investigation of human mobility patterns.

We characterize human movement patterns by five heavy-tailed distributions, i.e., power law, power law with cutoff, exponential, stretched exponential and lognormal distributions (Clauset, Shalizi and Newman 2009). In contrast to earlier studies that sought power law-like distributions only, we believe that all the five models can characterize human movement patterns. Thus, we make a pair-wise comparison among the five models and choose one best fit to characterize human movement patterns. Another distinguished feature of this study is that we placed considerable effort into data processing to extract the geometric and topological properties for exploring human mobility patterns. The basic principle of the data processing is the head/tail division rule that was formulated for empirical data that have a heavy-tailed distribution (Jiang and Liu 2011). We processed over 4 million captured locations of US domestic flights within an 11-day time period. From the massive amount of locations or points $(x, y, z, t)$, we extracted over 32,000 routes and 200,000 flights and established a topological relationship for over 700 airports across the country and within the specified time period (see the next section on data and data processing). We designated the flight movement as a proxy for human movement at the country level and examined the patterns from both geometric and topological perspectives. To further explore human mobility patterns, we setup various agent-based simulations to uncover the underlying mechanisms.

## 2. Data and data processing

We obtained, in total, 7,685,948 flight locations or points within 11 days between 8 and 18 August 2010, i.e., $(x_i, y_i, z_i, t_i)$ where $i = 7,685,948$. The starting time on the 8[th] was 1:36AM, while the ending time on the 18[th] was 8:29AM. The locations or points are referenced to the World Geodetic System, 1984. Apart from the location information, there are some other attributes, such as airline code, and flight number and speed; see Table 1 for a small sample.

Table 1: A sample of original flight data captured by GPS receivers

| Airline | Flight | Speed | Altitude | Latitude | Longitude | Timestamp |
|---|---|---|---|---|---|---|
| MES | 3449 | NULL | NULL | 19.221348 | 90.215218 | 08-09-2010 01:20:19 |
| COM | 321 | 323 | 8100 | 35.346111 | -89.633331 | 08-08-2010 23:49:26 |
| …… | | | | | | |
| MES | 3449 | NULL | NULL | 19.221348 | 90.215218 | 08-09-2010 01:20:19 |
| COM | 321 | 251 | 2900 | 35.165279 | -89.98111 | 08-08-2010 23:54:26 |
| …… | | | | | | |
| MES | 3449 | 134 | 27000 | NULL | NULL | 08-09-2010 02:03:44 |
| COM | 321 | 173 | 34000 | 34.960835 | -89.987221 | 08-08-2010 23:59:16 |

We had to process the data, initially a massive set of points (500MB), to determine the individual airports and how the airports are linked or related by flights or routes. We extracted two types of graphs: route graphs and flight graphs. The difference between a route graph and a flight graph is that the former bears a binary relationship while the latter is weighted by the number of flights. In other words, a route has at least one flight; there would be no route without a single flight. To build up the two graphs, first we extracted the individual flights. This process is shown in Figure 1 and is described in details as follows.

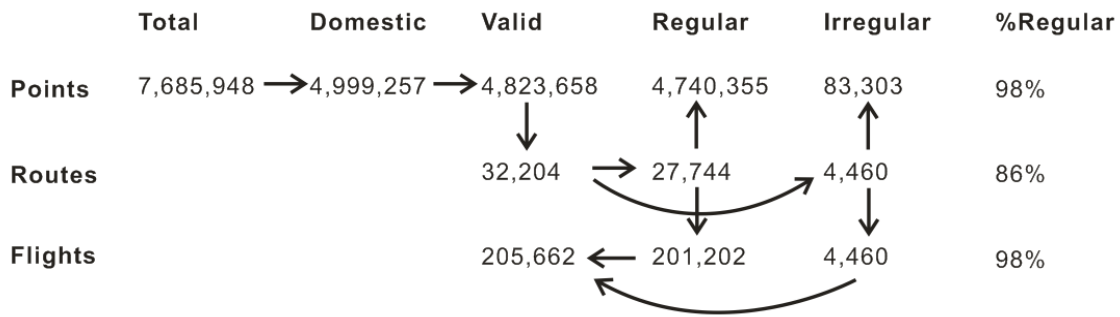

Figure 1: Flowchart of the data processing to extract 205,662 valid flights from 7,685,948 flight locations

Because we were interested in US domestic flights, we kept 4,999,257 domestic points with the tag 0 and filtered out international flights with the tag 1. There were thus $175,599 = 4,999,257 - 4,823,658$ invalid points. A point is invalid if there is the same point at the same instant in time or if one of the four dimensions is set to NULL. The 4,823,659 valid points belong to 32,204 valid routes that can be determined by a unique combination of airline code and flight number. The 32,204 routes are further divided into two types: regular and irregular.

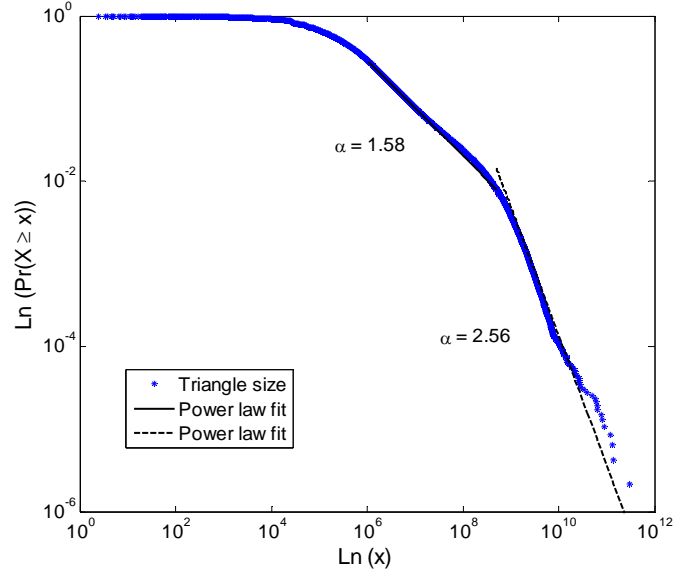

Figure 2: (Color online) Bipartite power law distribution of triangle sizes of the TIN

To distinguish between regular and irregular routes, we explored the time difference $\tau_i = t_i - t_{i-1}$ between two consecutive times $(t_{i-1}, t_i)$. For any irregular route that appears only once within the 11 days, $\tau_i$ does not change much. It fluctuates at approximately 5 minutes, which is the time interval to capture flight locations. Because all the points were captured while the flights were in service, for any regular route, $\tau_i$ fluctuated approximately 5 minutes while in service but around a larger value (e.g., six hours between this flight and the next one) while not in service. Given the difference, the arithmetic mean of $\tau_i$ for any irregular route ($\mu_2$) is greater than the standard deviation ($\sigma_2$), i.e, $\mu_2 > \sigma_2$, while the arithmetic mean of $\tau_i$ for any regular route ($\mu_1$) is far less than the standard deviation ($\sigma_1$), i.e., $\mu_1 << \sigma_1$. Using this statistical property, i.e., the difference between the arithmetic mean and standard deviation, we obtained 27,744 regular routes and 4,460 irregular routes. Consequently, the 27,744 regular routes contained 4,740,355 regular points, while the 4,460 irregular routes contained 83,303 irregular points. Thus, the percentages of regular points, regular routes and regular flights are pretty high respectively 98%, 86% and 98% as indicated in Figure 1.

Note that one irregular route is considered to be one irregular flight, i.e., the 4,460 irregular routes equal the same number of flights. However, one regular route contains multiple flights (Figure 3a).The process of deriving multiple flights from a route is achieved through the arithmetic mean of $\tau_i$ and the standard deviation $\sigma_i$ within each route. Those points with $\tau_i$ less than the sum of the mean and the standard deviation belong to a complete flight with the starting point as the origin and ending point as the destination (c.f., Figure 3b for an illustration). The 27,744 regular routes in fact contained 201,202 repeated flights. Eventually, we obtained 205,662 valid flights by adding the 4,460 irregular flights to the 201,202 regular flights.

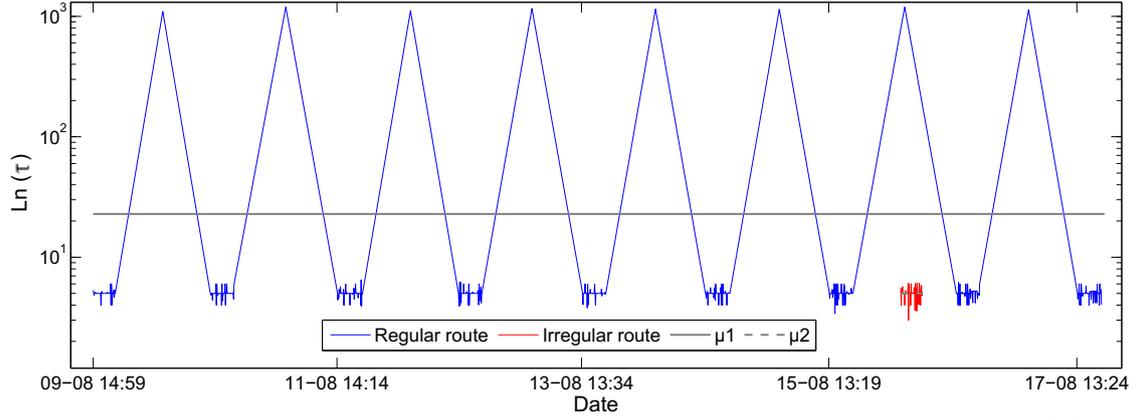

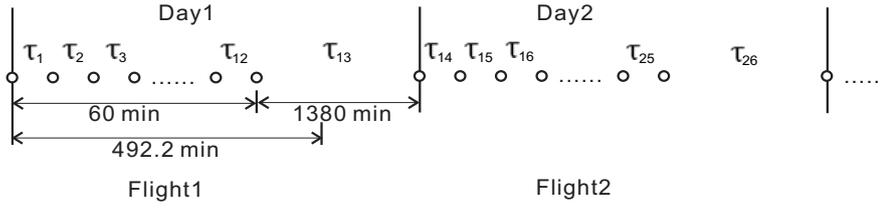

Figure 3: Derivation of multiple flights from one route: (a) One irregular route (red) and one regular route including 9 regular flights (blue), and (b) a fictive example of daily flights with the time intervals being precisely 5 minutes while in service

(Note: We adopt a fictive example to illustrate how to derive multiple flights from a regular route, i.e, how to derive those flight locations as complete flights between the origin and the destination. For the sake of simplicity, we assume a daily flight between two airports lasting one hour. Every day there would be 13 time intervals ($\tau_i$, where $i = 1,2,...13$), $\tau_i = 5$ for $1 \leq i \leq 12$, but $\tau_i = 23 \times 60$ for $i = 13$ in which period a flight is not in service. The arithmetic mean of all time intervals is $(5 \times 12 + 23 \times 60)/13 = 110.8$ minutes, and the standard deviation is 381.4. Given the fact that the time intervals are not precisely 5 minutes, we can use the sum of the arithmetic mean and the standard deviation (110.8 + 381.4) = 492.2 to obtain the consective points as a complete flight journey.)

With the 205,662 valid flights (or more specifically, the corresponding regular and irregular points as shown in Figure 1), we could extract the individual airports for setting up the route graphs and flight graphs. Every flight contains an origin (O) and a destination (D), so in theory there are $411,324 = 205,662 \times 2$ O/D locations. However, we retrieved only 410,536 O/D locations, slightly less than the theoretical estimate, because some O/D locations are shared. From the 410,536 points, we generated four triangulated irregular networks (TIN) for four areas: the Mainland, Alaska, Hawaii, and Puerto Rico and the Virgin islands. The TIN triangle sizes follow a bipartite power law distribution (Figure 2). Consequently, according to the head/tail division rule (Jiang and Liu 2011) and using the arithmetic mean of the TIN triangle sizes, we placed the triangles into two categories: those smaller than the mean and those greater than the mean. We designated those smaller than the mean as the airports, since airports are highly crowded with O/D points. Eventually, we generated 732 airports and determined the airport topologies in terms of both flights and routes for further analysis (Figure 4).

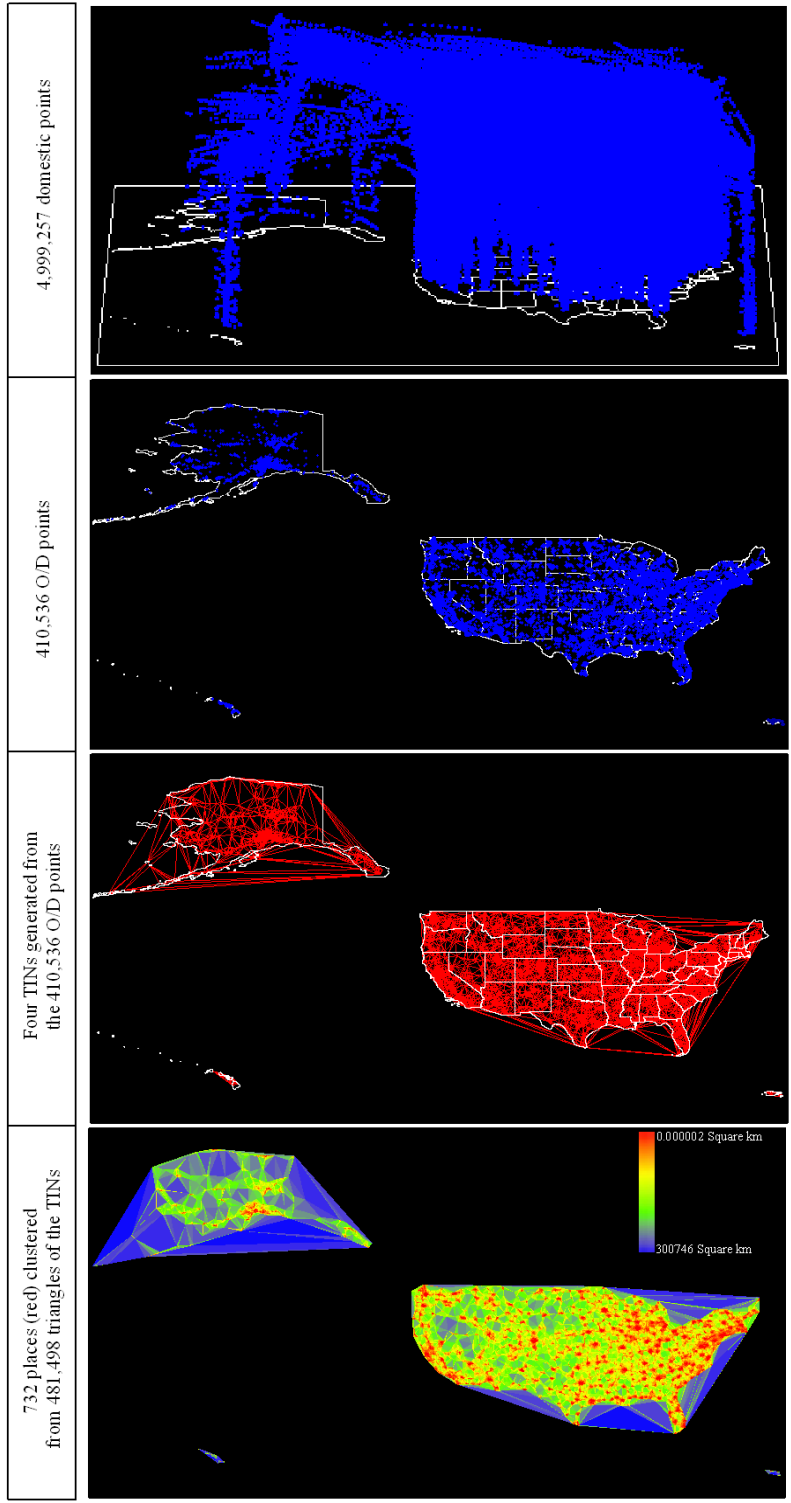

Figure 4: (Color online) Illustration of data processing from a point cloud to 732 natural airports

## 3. Mathematical characterization of flights or human mobility patterns

The central theme of our study is to determine which heavy tailed distribution characterizes human mobility patterns. The heavy tailed distributions refer often to three basic nonlinear relationships between quantity x and its probability: power law, lognormal and exponential, and their degenerate versions: power law with an exponential cutoff and stretched exponential (Clauset, Shalizi and Newman 2009). The focus of the work by Clauset and his colleagues was to detect a power law distribution and differentiate it from the four alternatives based on some statistical tests. Their methods can be summarized by three steps: (1) compute $x_{min}$ and $\alpha$ for a power law fit, (2) conduct Kolmogorov-Smirnov (KS) test and calculate the p value to assess the goodness of fit, and (3) compute the likelihood ratio (LR) values to compare with alternatives in case the fit cannot pass the KS test. In this paper, we adopt similar methods but do so to determine any heavy tailed distribution and differentiate it from the alternatives (see Appendix A). Let us first introduce the five distributions.

The power law distribution indicates a nonlinear relationship between a quantity (x) and its probability (y), given by $y = x^{-\alpha}$ in its simplest form, where $\alpha$ is usually between 1 and 3. This is an idealized expression. In real world data, the power law relationship appears only for the tail while x is greater than a minimum x. The power law is expressed as follows:

$$y = C_1 x^{-\alpha} \qquad , \qquad [1]$$

$$\text{where } C_1 \text{ is } (\alpha-1)x_{min}^{\alpha-1} \qquad [2]$$

The power law has a degenerate form, i.e., a power law with an exponential cutoff. In other words, the tail is not a perfect power law distribution but rather a mixture of power law and exponential.

$$y = C_2 x^{-\alpha} e^{-\lambda x} \qquad , \qquad [3]$$

$$\text{where } C_2 \text{ is } \frac{\lambda^{1-\alpha}}{\Gamma(1-\alpha, \lambda x_{min})} \qquad [4]$$

In its simplest format, an exponential relationship between x and y is expressed by $y = e^x$. Taking the logarithm of both sides of the expression, we obtain $\ln y = x$, which implies that the relationship between x and $\ln y$ is linear. More generally, an exponential distribution is expressed as follows:

$$y = C_3 e^{-\lambda x} \qquad , \qquad [5]$$

$$\text{where } C_3 \text{ is } \lambda e^{\lambda x_{min}} \qquad [6]$$

A degenerate version of the exponential function is the stretched exponential, i.e.,

$$y = C_4 x^{\beta-1} e^{-\lambda x^\beta} \qquad , \qquad [7]$$

$$\text{where } C_4 \text{ is } \beta\lambda e^{\lambda x_{min}^\beta} \qquad [8]$$

The last heavy tailed distribution is lognormal. Literally, lognormal implies that the logarithm of a quantity x follows a normal or Gaussian distribution, i.e.,

$$y = C_5 \frac{1}{x} e^{\left[\frac{(\ln x - \mu)^2}{2\sigma^2}\right]} \qquad , \qquad [9]$$

$$\text{where } C_5 \text{ is } \sqrt{\frac{2}{\pi\sigma^2}} \left[erfc\left(\frac{\ln x_{min} - \mu}{\sqrt{2}\sigma}\right)\right]^{-1} \qquad [10]$$

As shown in the above formulas, every one of the heavy tailed distributions consists of two parts: the main function that characterizes the distribution and the constant $C_i$. There is a series of parameters that describe the functions and constants, i.e., $x_{min}, \alpha, \beta, \lambda, \sigma, \mu$. The objective of Clauset, Shalizi and Newman (2009) was to detect a power law distribution in the form of equation [1] and to determine how to differentiate it from alternatives in the form of equations [3], [5], [7] and [9]. This paper, however, seeks a best fit function for characterizing various properties of human mobility patterns.

The methods for identifying the power law distributions can be extended for identifying any heavy tailed distribution. The methods can be used not only (1) to find the best fit models but also (2) to evaluate how good the fit is. For task (1), all of the parameters for the above models can be computed (see Appendix A). For task (2), there are two types of methods for the goodness-of-fit test. The first method is the KS test. For a real world dataset, there is a hypothesized model for $x \succ x_{min}$. The maximum difference ($\delta$) between the cumulative distribution function (CDF) of the dataset and the CDF of the hypothesized model indicates the goodness-of-fit or how close the model fits the real world dataset:

$$\delta = \max_{x \geq x_{min}} |f(x) - g(x)| \qquad , \qquad [11]$$

where $f(x)$ is the CDF of the dataset with a value of at least $x_{min}$ and $g(x)$ is the CDF for the hypothesized model that best fits the dataset for $x \geq x_{min}$.

However, to quantify the fit, we generated 1000 synthetic datasets according to the hypothesized model. The 1000 synthesized datasets had 1000 corresponding hypothesized models for $x \succ x_{min}$. Following the same idea of the maximum difference between the model and the data, we would have 1000 $\delta_i$, indicating the differences between the hypothesized models and the synthetic data. Eventually a goodness-of-fit index p was defined as a ratio of the number of $\delta_i$ whose values are greater than $\delta$ to 1000. If the p value is greater than a given threshold (e.g., 0.01), then the hypothesized model is acceptable, implying that 10 among the 1000 have a maximum distance greater than $\delta$.

For models that could not pass the above KS test, we needed alternative methods to compare the two competing models. One of the most frequently used methods is to calculate the likelihood ratio of two models. Because in this study, the KS test was sufficient to identity the potential models, we will not introduce the methods; interested readers can refer to Clauset, Shalizi and Newman (2009) for more details.

### 4. Simulations of human mobility in comparison with the observed
We simulated human movement using moving behaviors as shown in Figure 5 to uncover the underlying mechanism of human mobility patterns. One of the mechanisms is based on geometric distance (G), three of the mechanisms are based on topological properties (T1, T2, and T3), and a final one is a preferential return (PR) suggested by Song et al. (2010). For each scenario or mechanism, we setup 500 moving agents, moving in a route graph or flight graph from node to another for about 1000 times until the visited times are highly correlated to the degree of node connectivity (e.g. R square > 0.9). To this point, we think the simulation is saturated. Through the simulations, we obtained the simulated travel lengths (instead of traffic flow) and compared them with the observed ones. Note that the simulations are based on the topology of the 732 airports, both the route graphs and the flight graphs. The simulated travel lengths were was compared to the observed ones between the 732 airports. The comparison was based on a similar idea to the one illustrated in equation [11], i.e., the maximum distance between two CDF of the simulated and observed travel lengths. We ran each scenario 100 times and averaged the travel lengths for comparison to the observed ones.

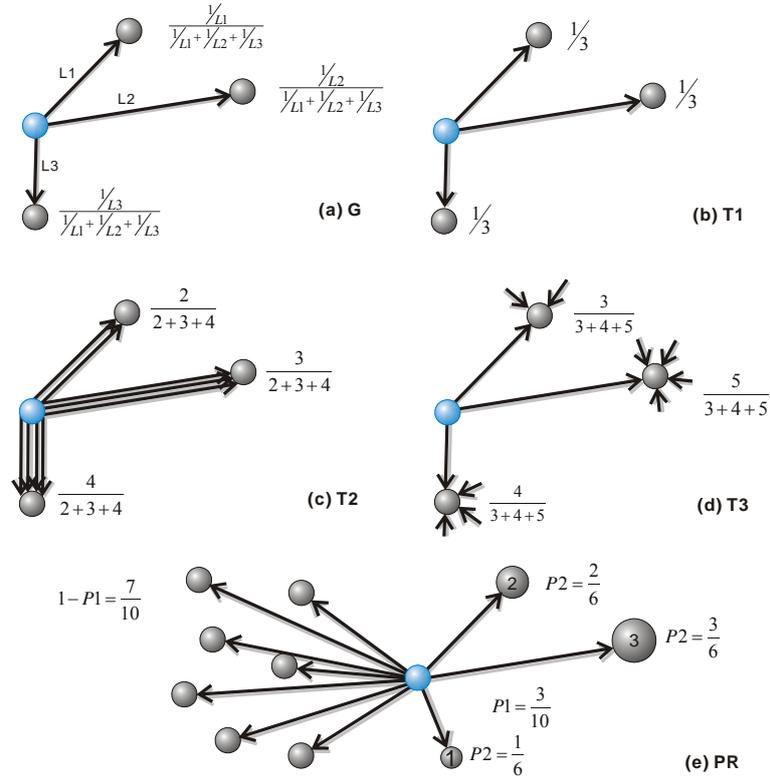

Figure 5: (Color online) Five scenarios of simulations indicated by (a) G = geometry (route graph), (b) T1 = topology 1 (route graph), (c) T2 = topology 2 (flight graph), (d) T3 = topology 3 (route graph), and (e) PR = preferential return (route graph)

(NOTE: A moving agent is currently at the blue node, and the agent wants to make the next move with a different probability as indicated. The geometric scenario is based on geometric distance, i.e., the closer, the better the probability (panel a). The next three behaviors are based on some topological properties: T1 is simply based on linkage (yes/no – whether or not there is a flight) (panel b), T2 considers the actual weight (how many actual flights) (panel c), while T3 consider how attractive its neighbors are (panel d). Preferential return is based on Barabasi's preferential attachment (Song et al. 2010), i.e., there is a higher chance to return to the nodes that were previously visited. In panel (e), all the linked nodes are placed into categories: the visited nodes to the right (the times are indicated by the numbers), and the unvisited nodes to the left (gray nodes). The first probability (P1) is according to the percentage of the two category nodes, while the second probability (P2) relates to a preference to return to those previously visited nodes.)

## 5. Results and discussion

We found that human mobility patterns are mainly shaped by the underling topology of airports and have little to do with human movement behaviors and distances. This finding is consistent with the recent studies (e.g., Jiang and Jia 2011, Han et al. 2011), but the experimental settings were rather different. For example, the two previous studies dealt with simulations: the former for street-constrained movement, the latter work for simulating movement in some idealized topology or networks. The current study is considered to be comprehensive because it covers (1) massive data processing, (2) mathematical characterization of the observed mobility patterns, and (3) agent-based simulations of different movement to compare with the observed flow or patterns. In what follows, we report the results related to those three main topics.

The detailed data processing is described in details in the above Section 2. A note-worthy result of the data processing is that the head/tail division rule (Jiang and Liu 2011) is applicable for delineating airports. The head/tail division rule states that *for given a variable X, if its values x follow a heavy-tailed distribution, then the mean (m) of the values can divide all the values into two parts: a high percentage in the tail and a low percentage in the head*. It should be noted that the heavy-tailed distribution is obtained by plotting ranking in a decreasing order as the x-axis, and the corresponding values as the y-axis, so-called rank-size plot or Zipf plot. The two parts, or both the head and the tail, correspond well to human cognition about real world phenomena, such as rich and poor, recession and prosperity, urban and rural, just to name a few examples. This rule was initially used to delineate city boundaries when it was first formulated. Herein, the application of the rule provides further evidence for its usefulness. Note that the airports are naturally identified depending on the density of O/D points – so-called natural airports. The natural airport boundaries may deviate from the real ones, but the boundaries are sufficiently accurate to set up airport relationships or topology.

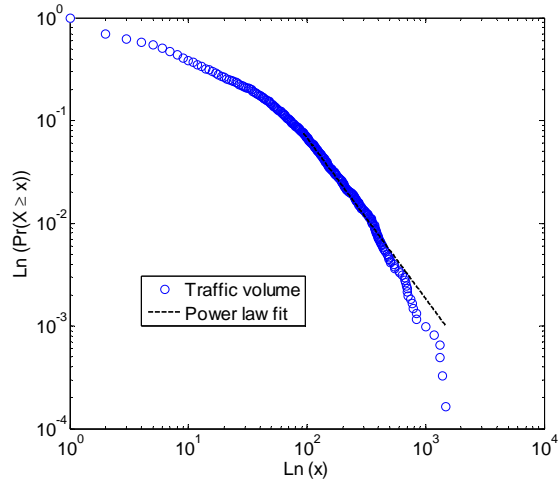

Figure 6: (Color online) Power law distribution of traffic volume

We found that topological properties, including traffic volume (flight numbers between airports) and degree of connectivity (number of flights arriving or departing from the individual airports), show a power law distribution. Figure 6 illustrates the power law distribution in a double logarithm plot: $y = 183 * x^{-2.56} (x \geq 93)$ with the KS test index p = 0.1. To further illustrate the hierarchy of traffic volume, we reduced it into different levels of detail according to the head/tail division rule (*c.f.*, Figure 7). Both the indegree and outdegree exhibit a striking power law distribution with slightly different parameter settings: $y = 20.68 * x^{-1.72} (x \geq 106)$ with the KS test index p = 0.02, and $y = 13.04 * x^{-1.67} (x \geq 84)$ with the KS test index p = 0.05 (Figure 8).

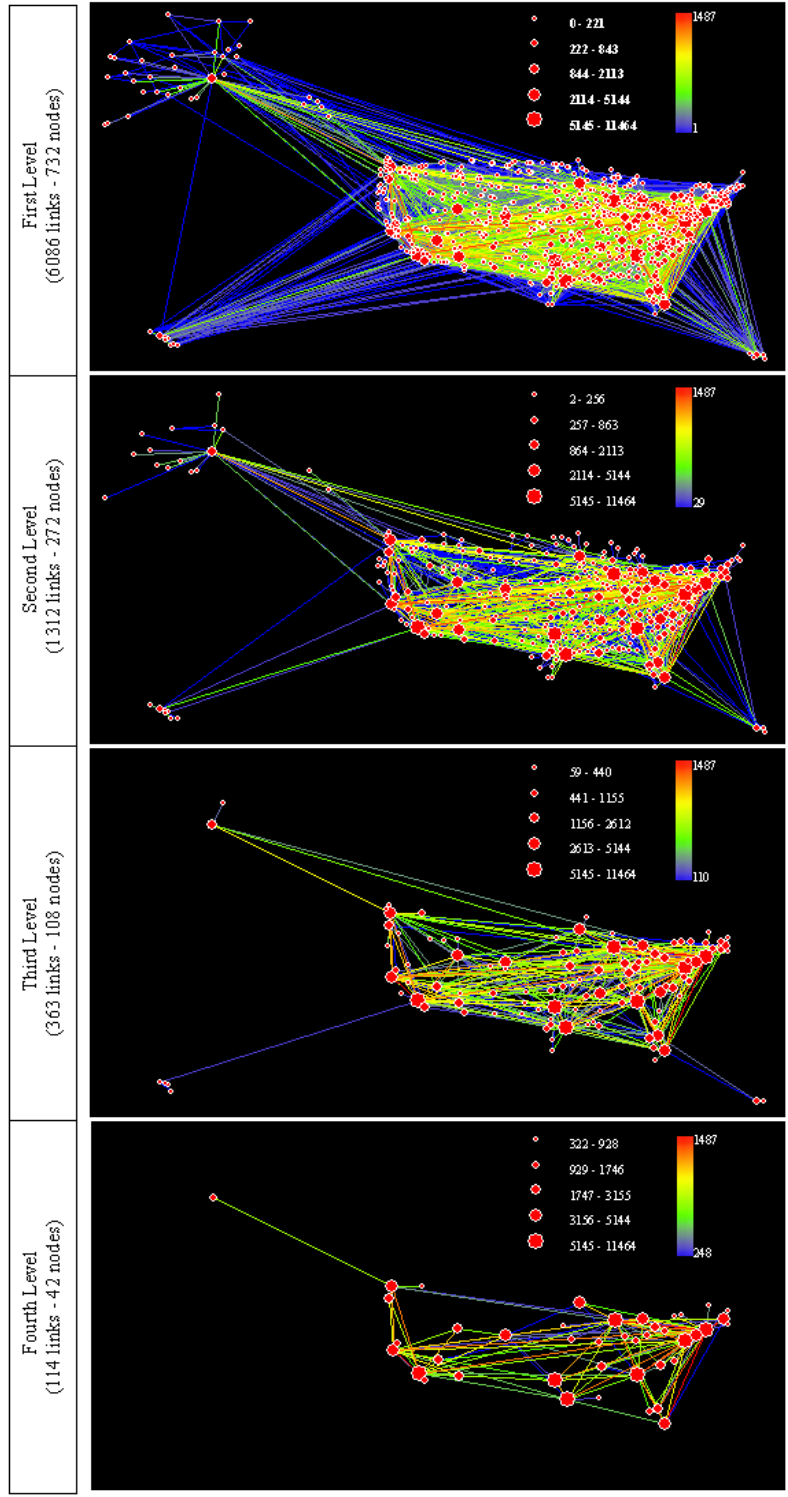

Figure 7: (Color online) Four levels of hierarchy of traffic volume

(NOTE: The first level represents the most detailed level, in which all the routes with at least one flight are shown together with the related airports or nodes. The second level is obtained from the first level by selecting those links with more than the mean number (29) of flights. This same procedure continues recursively for deriving the third and fourth levels. The sizes of the nodes indicate the magnitude of indegree, which is highly correlated with that of the outdegree.)

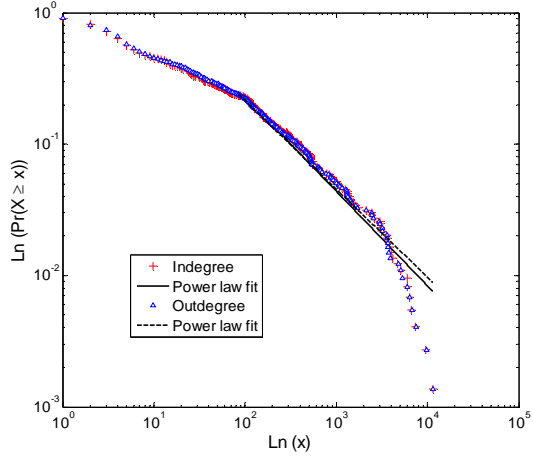

Figure 8: (Color online) Power law distributions of airport degrees

We found that human travel lengths demonstrate an exponential distribution with a power law cutoff, i.e.,

$$y = \begin{cases} 0.0012 \times e^{-0.0008x} & (x \leq 9000\ km) \quad p = 0.3 \\ 55845000 \times x^{-2.89} & (x > 9000\ km) \quad p = 0.5 \end{cases}$$

This result indicates that the distribution is exponential for travel lengths less than 9000 km, whereas it is a power law for travel lengths greater than 9000 km (Figure 9). It should be noted that the power law part contains only 128 flights (which can be ignored due to insufficient statistics), while the exponential part contains 205,534 flights. This result is very different from the early finding, which claims that human travel lengths demonstrate a power law with an exponential cutoff (Brockmann, Hufnage and Geisel 2006, Gonzalez, Hidalgo and Barabási 2008, Jiang, Yin and Zhao 2009). In other words, for those short lengths less than the threshold, there is a power law distribution, while for long ones, there is an exponential distribution. This conclusion is opposite from our finding.

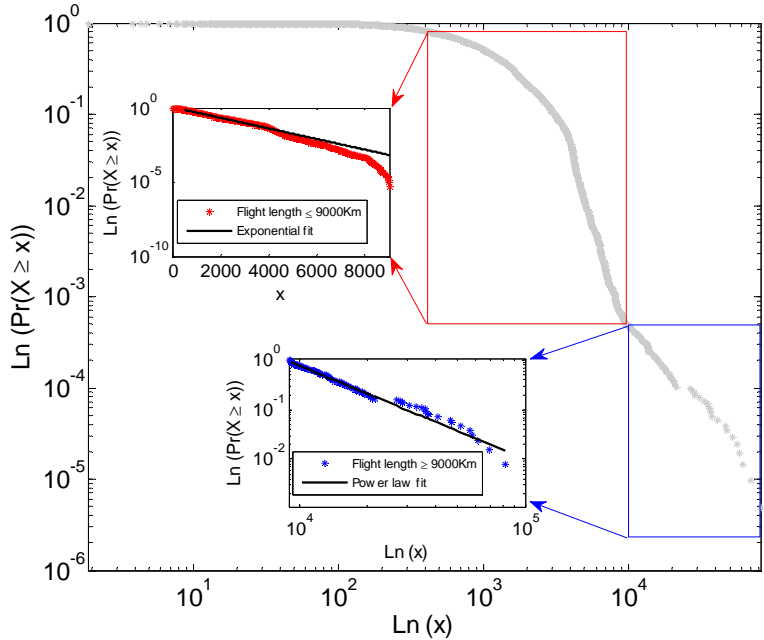

Figure 9: (Color online) Exponential distribution with a power law cutoff for travel length

Note that the above findings are based on a flight graph whose links are weighted by the number of flights. However, earlier studies are mainly based on a route graph that is a binary graph – the relation between airports is either 1 or 0 (e.g., Guimerà and Amaral 2004, Guimerà et al. 2005). In other words, there was no traffic volume information involved in the earlier studies. The airport topology used in the previous studies is purely topological, i.e., any two airports with a link as long as there is one flight regardless of the number of flights. Our airport topology is weighted, i.e., the link is weighted by the number of flights. A binary route graph cannot fully characterize human movement patterns or flight patterns. Nevertheless, we reduced the flight graph to a route graph and examined both the topological and geometric properties. We found that with the binary route graph both degree and betweenness centralities exhibit a power law distribution with an exponential cutoff. This result regarding the binary route graph is consistent with earlier studies (e.g., Guimerà and Amaral 2004, Guimerà et al. 2005). The route lengths demonstrate an exponential distribution, i.e., $y = 0.002 \times e^{-0.001x} (x \geq 768)$ with the KS test index p = 0.8 (Figure 10). If we take the route lengths as a proxy for travel lengths, then the result is also different from what was reported in the literature that claims a power law or power law with an exponential cutoff (Brockmann, Hufnage and Geisel 2006, Gonzalez, Hidalgo and Barabási 2008, Jiang, Yin and Zhao 2009). Another interesting finding is the rich club phenomenon observed with the binary route graph, i.e., the top 50 airports constitute a nearly complete graph with at least 41 routes to other airports.

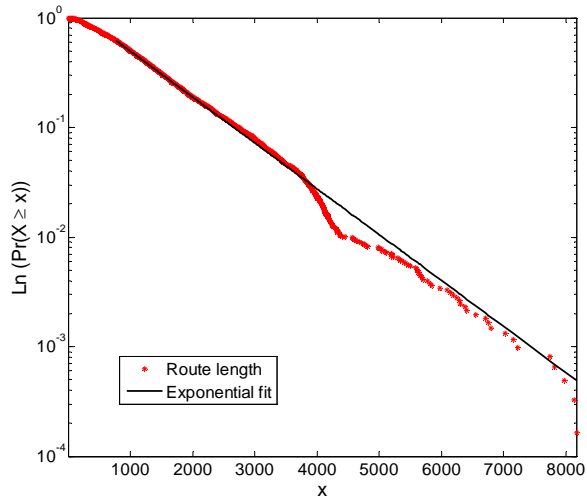

Figure 10: (Color online) Exponential distribution of route lengths

Through agent-based simulations of human mobility (*c.f.*, Section 4 and Figure 5), we found that the travel lengths with scenario T1, based on the route graph and random moving behavior, best match the observed travel lengths. T1 is sequentially followed by T2, PR, T3 and G in terms of matching degree. This result is counter-intuitive because we naively tend to believe that the other two topologically based scenarios T2 and T3 would have a better effect in replicating the observed travel lengths. Scenario T2 is based on the de facto flight graph, while scenario T3 considers how attractive the neighbors are. The scenario based on preferential return is good but not the best. This finding illustrates the underlying topological structure (the binary one rather than the weighted one), and random motion can describe the movement patterns well. In other words, human movement patterns have little to do with geometric factors or moving behavior.

## 6. Conclusion

Massive tracking data of flights provide a new means of studying human mobility patterns. In this paper, we assume US flights to be a proxy for human movement at the country level. From the massive location information of US flights within an 11-day period, we extracted individual routes, flights and airports and setup both flight graphs and route graphs for exploring the mobility patterns. The resulting data are published together with the paper. It is found that topological properties, such as traffic volume and airport degrees, are indeed power-law distributed. Travel lengths exhibit an exponential distribution with a power law cutoff rather than a power law distribution with an exponential cutoff as previous studies illustrated. To uncover the underlying mechanisms, we constructed various simulation scenarios based on both flight graphs and route graphs and found that route graphs with a random-motion behavior replicate the observed travel lengths pattern. This finding indicates that human movement patterns at a collective level are mainly shaped by the underlying structure – the simplest topology represented by the binary route graph.

**Acknowledgments**
The flight data was collected by Flytecomm Inc., and we are grateful for their permission to use it in this paper.

**References:**
Brockmann D., Hufnage L., and Geisel T. (2006), The scaling laws of human travel, *Nature*, 439, 462 – 465.
Clauset A., Shalizi C. R., and Newman M. E. J. (2009), Power-law distributions in empirical data, *SIAM Review*, 51, 661-703.
Gonzalez M., Hidalgo C. A., and Barabási A.-L. (2008), Understanding individual human mobility patterns, *Nature*, 453, 779 – 782.
Guimerà R. and Amaral L. A. N. (2004), Modeling the world-wide airport network, *The European Physical Journal B*, 38(2), 381-385.
Guimerà R., Mossa S., Turtschi A. and Amaral L. A. N. (2005), The worldwide air transportation network: Anomalous centrality, community structure, and cities' global roles, *Proceedings of the National Academy of Sciences of the United States of America*, 102(22), 7794-7799.
Hägerstrand T. (1970), What about people in regional science? *Papers of the Regional Science Association*, 24, 7 – 21.
Han X., Hao Q., Wang B. and Zhou T. (2011), Origin of the scaling law in human mobility: Hierarchy of traffic systems, *Physical Review E*, 83, 036117.
Hillier B., Penn A., Hanson J., Grajewski T. and Xu J. (1993), Natural movement: configuration and attraction in urban pedestrian movement, *Environment and Planning B: Planning and Design*, 20, 29-66.
Hufnagel L., Brockmann D. and Geisel T. (2004), Forecast and control of epidemics in a globalized world, *Proceedings of the National Academy of Sciences*, 101(42), 15124-15129.
Jiang B. and Jia T. (2011), Agent-based simulation of human movement shaped by the underlying street structure, *International Journal of Geographical Information Science*, 25(1), 51 – 64.
Jiang B. and Liu X. (2011), Scaling of geographic space from the perspective of city and field blocks and using volunteered geographic information, *International Journal of Geographical Information Science*, x, xx-xx, Preprint, arxiv.org/abs/1009.3635.
Jiang B., Yin J. and Zhao S. (2009), Characterizing human mobility patterns in a large street network, *Physical Review E*, 80, 021136.
Lee K., Hong S., Kim S. J., Rhee I. and Chong S. (2009), SLAW: A mobility model for human walks, *IEEE INFOCOM 2009*, IEEE.
Roth C., Kang S. M., Batty M, and Barthélemy M. (2011), Structure of urban movements: Polycentric activity and entangled hierarchical flows, *PLoS ONE*, 6(1), e15923.
Song C., Koren T., Wang P. and Barabási A. (2010), Modelling the scaling properties of human mobility, *Nature Physics*, DOI: 10.1038/NPHYS1760.
Ypma T. J. (1995), Historical development of the Newton-Raphson method, *SIAM Review*, 37(4), 531–551.

**Appendix A: Estimation of the parameters for some heavy tailed distributions based on the maximum likelihood method**

Based on the methods suggested by Clauset, Shalizi and Newman (2009) for detecting a power law distribution, in this appendix, we present various procedures for estimating the parameters of the other four heavy–tailed, continuous distributions. The primary method used is the maximum likelihood method.

1. Maximum likelihood estimation for the continuous exponential distribution.

The general form of the continuous exponential distribution can be described by

$$y = C_3 e^{-\lambda x} \qquad [1.1]$$

where $\lambda$ is the rate parameter ( $\lambda > 0$ ) and $C_3$ is the normalizing constant.

1.1 Calculating the normalizing constant $C_3$

It is obvious that $y$ diverges as $x \to 0$; thus, the above equation does not hold for all values of $x$. There must be a lower bound $x_{min}$ for the exponential distribution. Supposing that $x_{min}$ and $\lambda$ are known, we can easily derive the normalizing constant $C_3$.

To find the normalizing constant, we use the fact that

$$\int_{x_{min}}^{+\infty} C_3 e^{-\lambda x} dx = 1 \qquad [1.2]$$

Thus,

$$\frac{-C_3}{\lambda} \int_{x_{min}}^{+\infty} e^{-\lambda x} d(-\lambda x) = 1 \qquad [1.3]$$

$$\frac{-C_3}{\lambda} e^{-\lambda x} \bigg|_{x_{min}}^{+\infty} = 1 \qquad [1.4]$$

$$\frac{C_3}{\lambda} e^{-\lambda x_{min}} = 1 \qquad [1.5]$$

Finally,

$$C_3 = \lambda e^{\lambda x_{min}} \qquad [1.6]$$

1.2 Estimating the rate parameter $\lambda$

Now suppose, given an empirical dataset containing n observations $x_i (i = 1, 2, ..., n)$, we want to know the probability that this dataset is drawn from the continuous exponential distribution model. This probability is also known as the likelihood, and it is described by

$$L = \prod_{i=1}^{n} y_i \qquad [1.7]$$

Thus, in this case,

$$L(\lambda) = (\lambda e^{\lambda x_{min}})^n \prod_{i=1}^{n} e^{-\lambda x_i} \qquad . \qquad [1.8]$$

We now wish to find the value of $\lambda$ that maximizes $L(\lambda)$. Mathematically, this value can be obtained by setting the derivative $dL(\lambda)/d\lambda$ equal to 0 and solving for $\lambda$. However, it is tedious to find the derivative of $L(\lambda)$ because it is a product of functions in $\lambda$. Hence, we adopt the natural logarithm form of $L(\lambda)$ because it is easier to find the value of $\lambda$ that maximizes $Ln[L(\lambda)]$. We have

$$Ln[L(\lambda)] = nLn\lambda + n\lambda x_{min} - \lambda \sum_{i=1}^{n} x_i \qquad [1.9]$$

The derivative of $Ln[L(\lambda)]$ with respect to $\lambda$ is

$$\frac{dLn[L(\lambda)]}{d\lambda} = \frac{n}{\lambda} + nx_{min} - \sum_{i=1}^{n} x_i \qquad [1.10]$$

Thus, the value of $\lambda$ that maximizes $Ln[L(\lambda)]$ is the solution of the equation

$$\frac{n}{\lambda} + nx_{min} - \sum_{i=1}^{n} x_i = 0 \qquad [1.11]$$

Solving, we obtain the estimates $\hat{\lambda}$ as

$$\hat{\lambda} = n\left[\sum_{i=1}^{n} x_i - nx_{min}\right]^{-1} \qquad [1.12]$$

2. Maximum likelihood estimation for the continuous stretched exponential distribution.

The general form of the continuous stretched exponential distribution can be described as:

$$y = C_4 x^{\beta-1} e^{-\lambda x^{\beta}} \qquad [2.1]$$

where $\lambda$ is the rate parameter ($\lambda > 0$), $\beta$ is the stretching exponent ($\beta > 0$) and $C_4$ is the normalizing constant. For $\beta = 1$, it degenerates to an exponential distribution as discussed above.

2.1 Calculate the normalizing constant $C_4$

To calculate the normalizing constant $C_4$, we impose a lower bound $x_{min}$ for the stretched exponential distribution. Let us assume that $x_{min}$, $\beta$ and $\lambda$ are known; we can easily derive the normalizing constant $C_4$.

To find the normalizing constant, we use the fact that

$$\int_{x_{min}}^{+\infty} C_4 x^{\beta-1} e^{-\lambda x^{\beta}} dx = 1 \qquad , \qquad [2.2]$$

Thus,

$$\frac{-C_4}{\lambda\beta} \int_{x_{min}}^{+\infty} e^{-\lambda x^{\beta}} d(-\lambda x^{\beta}) = 1 \qquad [2.3]$$

$$\frac{-C_4}{\lambda\beta} e^{-\lambda x^{\beta}} \Big|_{x_{min}}^{+\infty} = 1 \qquad [2.4]$$

$$\frac{C_4}{\lambda\beta} e^{-\lambda x_{min}^{\beta}} = 1 \qquad , \qquad [2.5]$$

Finally,

$$C_4 = \lambda\beta e^{\lambda x_{min}^{\beta}} \qquad , \qquad [2.6]$$

2.2 Estimating the rate parameter $\lambda$ and stretching exponent $\beta$

Now suppose, given an empirical dataset containing n observations $x_i (i = 1, 2, ..., n)$, we want to know the likelihood that this dataset is drawn from the continuous stretched exponential distribution model. In this case, the likelihood is given by

$$L(\lambda, \beta) = \prod_{i=1}^{n} \beta \lambda x_i^{\beta-1} e^{\lambda(x_{min}^{\beta} - x_i^{\beta})} \quad , \qquad [2.7]$$

Furthermore,

$$Ln[L(\lambda, \beta)] = nLn\lambda + nLn\beta + (\beta - 1)\sum_{i=1}^{n} Lnx_i + n\lambda x_{min}^{\beta} - \lambda \sum_{i=1}^{n} x_i^{\beta} \quad . \quad [2.8]$$

The maximum likelihood estimators of $\lambda$ and $\beta$ are the values that maximize $Ln[L(\lambda, \beta)]$. Taking derivatives with respect to both $\lambda$ and $\beta$, we obtain

$$\frac{\partial(Ln(L(\lambda, \beta)))}{\partial \lambda} = \frac{n}{\lambda} + nx_{min}^{\beta} - \sum_{i=1}^{n} x_i^{\beta} \qquad [2.9]$$

and

$$\frac{\partial(Ln(L(\lambda, \beta)))}{\partial \beta} = \frac{n}{\beta} + \sum_{i=1}^{n} Lnx_i + n\lambda x_{min}^{\beta} Lnx_{min} - \lambda \sum_{i=1}^{n} x_i^{\beta} Lnx_i \quad . \quad [2.10]$$

Setting the above two derivatives equal to 0, we obtain the following system of equations,

$$\begin{cases} \frac{n}{\lambda} + nx_{min}^{\beta} - \sum_{i=1}^{n} x_i^{\beta} = 0 \\ \frac{n}{\beta} + \sum_{i=1}^{n} Lnx_i + n\lambda x_{min}^{\beta} Lnx_{min} - \lambda \sum_{i=1}^{n} x_i^{\beta} Lnx_i = 0 \end{cases} \quad . \quad [2.11]$$

Solving the first equation, we obtain

$$\hat{\lambda} = n[\sum_{i=1}^{n} x_i^{\hat{\beta}} - nx_{min}^{\hat{\beta}}]^{-1} \quad . \quad [2.12]$$

Substituting $\hat{\lambda}$ for $\lambda$ in the second equation, we obtain

$$\frac{n}{\hat{\beta}} + \sum_{i=1}^{n} Lnx_i + n[\sum_{i=1}^{n} x_i^{\hat{\beta}} - nx_{min}^{\hat{\beta}}]^{-1} (nx_{min}^{\hat{\beta}} Lnx_{min} - \sum_{i=1}^{n} x_i^{\hat{\beta}} Lnx_i) = 0 \quad . \quad [2.13]$$

Notice that the left side of the above equation is known as $f(\hat{\beta})$, and our goal is to find the value of $\hat{\beta}$ that makes this function equal to 0. However, it is difficult to solve it directly to obtain the value of $\hat{\beta}$ because it is difficult to derive the inverse function of $f(\hat{\beta})$ as $\hat{\beta} = f^{-1}(x_i, x_{min})$. Hence, we use the simple Newton–Raphson method (Ypma, 1995) to derive the solution to $f(\hat{\beta})$. Once the value of $\hat{\beta}$ is determined, the value of $\hat{\lambda}$ is also determined.

The Newton–Raphson method is used for finding successively better approximations to the roots of a function $f$. Initially, we give an estimation of the root $x0$ that is reasonably close to the true root. Then, we draw a tangent line of the function $f$ at this estimation point, and the intersection point $x1$ between this tangent line and the x-axis is determined as the second estimation of the root, which is considered to be a better approximation. After several iterations, if the current estimation satisfies our accuracy requirement, then this process is terminated; otherwise, this process is repeated until a reasonable accuracy is accepted. Note that the derivative of the function $f$ should be calculated directly because the tangent line must be constructed at the estimation point. The following figure illustrates this method.

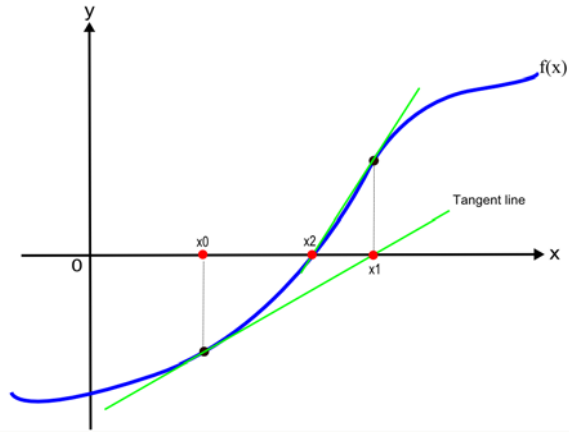

Figure A1: (Color online) Illustration of the Newton–Raphson method

Based on this method, firstly, we set the initial value of $\hat{\beta}$ as the exponent value of a pure power law distribution fitted by the observed dataset. Then, we obtain the derivative of this function,

$$f'(\hat{\beta}) = \frac{-n}{\hat{\beta}^2} + n \frac{[nx_{min}^{\hat{\beta}} Lnx_{min}^2 - \sum_{i=1}^{n} x_i^{\hat{\beta}} Lnx_i^2](\sum_{i=1}^{n} x_i^{\hat{\beta}} - nx_{min}^{\hat{\beta}}) + [nx_{min}^{\hat{\beta}} Lnx_{min} - \sum_{i=1}^{n} x_i^{\hat{\beta}} Lnx_i]^2}{(-nx_{min}^{\hat{\beta}})^2}$$

[2.14]

Next, following the steps of Newton–Raphson method, we can calculate the next better estimation value of $\hat{\beta}_e$ with the formula

$$\hat{\beta}_e = \hat{\beta} - \frac{f(\hat{\beta})}{f'(\hat{\beta})} .$$

[2.15]

Finally, this procedure will be terminated if

$$f(\hat{\beta}_e) < \delta ,$$

[2.16]

where $\delta$ is the acceptable error.

Otherwise, this procedure must be iterated until this condition is satisfied.

3. Maximum likelihood estimation for the continuous lognormal distribution.

The general form of the continuous lognormal distribution can be described by

$$y = (C_5/(x\sigma\sqrt{2\pi}))\exp(-(Lnx-\mu)^2/2\sigma^2) ,$$

[3.1]

where $\mu$ is the mean value, $\sigma$ is the standard deviation value and $C_5$ is the normalizing constant.

## 3.1 Calculating the normalizing constant $C_5$

To calculate the normalizing constant $C_5$, we impose a lower bound $x_{min}$ on the lognormal distribution. Let us assume that $x_{min}$, $\mu$ and $\sigma$ are known. Then, we can easily derive the normalizing constant $C_5$.

To find the normalizing constant, we use the fact that

$$\int_{x_{min}}^{+\infty} (C_5/(x\sigma\sqrt{2\pi})) \exp(-(Lnx-\mu)^2/2\sigma^2) dx = 1 \qquad . \qquad [3.2]$$

Thus,

$$\sqrt{2}\sigma \frac{C_5}{\sigma\sqrt{2\pi}} \int_{x_{min}}^{+\infty} \exp(-(Lnx-\mu)^2/2\sigma^2) d(\frac{Lnx-\mu}{\sqrt{2}\sigma}) = 1 \qquad [3.3]$$

$$\frac{C_5\sqrt{\pi}}{\sqrt{\pi}2} erfc(\frac{Lnx_{min}-\mu}{\sqrt{2}\sigma}) = 1 \qquad [3.4]$$

$$\frac{C_5}{2} erfc(\frac{Lnx_{min}-\mu}{\sqrt{2}\sigma}) = 1 \qquad . \qquad [3.5]$$

Finally,

$$C_5 = 2[erfc(\frac{Lnx_{min}-\mu}{\sqrt{2}\sigma})]^{-1} \qquad , \qquad [3.6]$$

where

$$erfc(x) = \frac{2}{\sqrt{\pi}} \int_x^{+\infty} e^{-t^2} dt \qquad . \qquad [3.7]$$

## 3.2 Estimating the mean parameter $\mu$ and the standard deviation parameter exponent $\sigma$

Suppose that, given an empirical dataset containing n observations $x_i (i=1,2,...,n)$, we want to know the likelihood that this dataset is drawn from the continuous lognormal distribution model. In this case, the likelihood is given by

$$L(\mu,\sigma) = \prod_{i=1}^n \frac{1}{x_i} \exp(-\frac{(Lnx_i-\mu)^2}{2\sigma^2}) [\sigma\sqrt{2} \int_{\frac{Lnx_{min}-\mu}{\sigma\sqrt{2}}}^{+\infty} e^{-t^2} dt]^{-1} \qquad , \qquad [3.8]$$

Furthermore,

$$Ln[L(\mu,\sigma)] = \sum_{i=1}^n -\frac{(Lnx_i-\mu)^2}{2\sigma^2} - \sum_{i=1}^n Lnx_i - nLn\sigma - nLn(\sqrt{2} \int_{\frac{Lnx_{min}-\mu}{\sigma\sqrt{2}}}^{+\infty} e^{-t^2} dt)$$

[3.9]

The maximum likelihood estimators of $\mu$ and $\sigma$ are the values that maximize $Ln[L(\mu,\sigma)]$. Here, we adopt the technique provided by the R package to obtain the estimation value of $\mu$ and $\sigma$ because it is not convenient to use the same strategy as in the case of the stretched exponential distribution to handle the above complicated log likelihood function.

R is a collection of powerful packages that is used for data manipulation, calculation and graphical display. It can be used as a statistical system that implements a rich set of classical and modern statistical techniques. The problem here of obtaining an estimate for the value of $\mu$ and $\sigma$ to maximize $Ln[L(\mu,\sigma)]$ can be treated as a problem of nonlinear optimization. In R, the function nlm() is used to carry out a minimization of the objective function f based on a Newton-type algorithm. Therefore, our problem can be solved as a minimization of the negative $Ln[L(\mu,\sigma)]$, and the procedure is demonstrated with the following pseudo-code.

```
Function LognormalFit (x, threshold)
   Set x=x (x≥threshold)
   Guess the initial value of both μ and σ using μ = Mean(Ln(x)) and σ = Std(Ln(x))
   Set estimation = nlm(f = -Ln[L(μ,σ)], p = (μ,σ))
   Return estimation
End Function
```

where the variable estimation returns the best estimation value for both $\mu$ and $\sigma$.

4. Maximum likelihood estimation for the continuous power law distribution with exponential cutoff.

The general form of the continuous lognormal distribution can be described by

$$y = C_2 x^{-\alpha} e^{-\gamma x} \quad , \quad [4.1]$$

where $\alpha$ is the power law exponent parameter, $\gamma$ is the exponential rate parameter and $C_2$ is the normalizing constant.

4.1 Calculating the normalizing constant $C_2$

To calculate the normalizing constant $C_2$, we impose a lower bound $x_{min}$ on the power law distribution with an exponential cutoff. Let us assume that $x_{min}$, $\alpha$ and $\gamma$ are known. Then, we can easily derive the constant $C_2$.

To find the normalizing constant, we use the fact that

$$\int_{x_{min}}^{+\infty} C_2 x^{-\alpha} e^{-\gamma x} dx = 1 \quad . \quad [4.2]$$

Thus,

$$\int_{x_{min}}^{+\infty} C_2 \gamma^{\alpha-1} (\gamma x)^{-\alpha} \exp(-\gamma x) d(\gamma x) = 1 \quad . \quad [4.3]$$

Substituting $t$ for $\gamma x$,

$$C_2 \gamma^{\alpha-1} \int_{x_{min}}^{+\infty} t^{-\alpha} \exp(-t) dt = 1 \quad . \quad [4.4]$$

Substituting $1-\alpha$ for $a$,

$$C_2 \gamma^{-a} \int_{\gamma x_{min}}^{+\infty} t^{a-1} \exp(-t) dt = 1 \quad . \quad [4.5]$$

Then,

$$C_2 \gamma^{-a} [\Gamma(a) - \Gamma(a)\Gamma(a, \gamma x_{min})] = 1 \quad . \quad [4.6]$$

Again, substituting $a$ for $1-\alpha$, and finally,

$$c = \frac{\gamma^{1-\alpha}}{\Gamma(1-\alpha)[1-\Gamma(1-\alpha,\gamma x_{\min})]} \quad , \quad [4.7]$$

where

$$\Gamma(1-\alpha) = \int_0^{+\infty} e^{-t} t^{-\alpha} dt \quad , \quad [4.8]$$

and

$$\Gamma(1-\alpha,\gamma x_{\min}) = \frac{\int_0^{\gamma x_{\min}} e^{-t} t^{-\alpha} dt}{\Gamma(1-\alpha)} \quad . \quad [4.9]$$

4.2 Estimating the power law exponent parameter $\alpha$ and the exponential rate parameter $\gamma$

Suppose that, given an empirical dataset containing n observations $x_i (i=1,2,...,n)$, we want to know the likelihood that this dataset is drawn from the continuous power law distribution with an exponential cutoff model. In this case, the likelihood is given by

$$L(\alpha,\gamma) = \prod_{i=1}^{n} \frac{\gamma^{1-\alpha}}{\Gamma(1-\alpha)[1-\Gamma(1-\alpha,\gamma x_{\min})]} x_i^{-\alpha} e^{-\gamma x_i} \quad . \quad [4.10]$$

Furthermore,

$$Ln[L(\alpha,\gamma)] = n(1-\alpha)Ln\gamma - nLn[\Gamma(1-\alpha)[1-\Gamma(1-\alpha,\gamma x_{\min})]] - \alpha \sum_{i=1}^{n} Lnx_i - \gamma \sum_{i=1}^{n} x_i$$

[4.11]

The maximum likelihood estimators of $\alpha$ and $\gamma$ are the values that maximize $Ln[L(\alpha,\gamma)]$. However, because of the complexity of the above log likelihood function, we adopt the same strategy as in the case of the lognormal distribution. One small difference is that here we use the function ConstrOptim() in R to carry out a minimization of the objective function f with the constraints on the range of parameters. Therefore, our problem can be solved as a minimization of the negative $Ln[L(\alpha,\gamma)]$ with parameter constraints, and the procedure is demonstrated with the following pseudo-code.

```
Function PowerlawExpFit (x, threshold)
   Set x = x (x≥threshold)
   Guess the initial value of α by fitting x to a pure power law distribution
   Guess the initial value of γ by fitting x to a pure exponential distribution
   Set the constraints as α≥-1 and γ≥0
   Set estimation = ConstrOptim(f = -Ln[L(α,γ)], p = (α,γ), constraints)
   Return estimation
End Function
```

Where variable estimation returns the best estimation value for both $\alpha$ and $\gamma$.